\documentstyle[prb,aps,epsfig]{revtex}

\begin{document}

\draft

\twocolumn[\hsize\textwidth\columnwidth\hsize\csname
@twocolumnfalse\endcsname
 
\title{Spectral properties of the one-dimensional two-channel Kondo 
 lattice model}

\author {A. M. Tsvelik}

\address {Department of Physics, University of Oxford, 
1 Keble Road, Oxford,  OX1 3NP, U.K.}

\author {C. I. Ventura~$^*$}

\address {Centro At\'omico Bariloche, 8400-Bariloche, Argentina.}

\maketitle

\begin{abstract}

{\small 

We have studied the energy spectrum of a one-dimensional Kondo lattice, 
where the localized magnetic moments have SU(N) symmetry and two channels of 
conduction electrons are present. At half filling, the system is shown 
to exist in two phases: one dominated by RKKY-exchange interaction effects, 
and the other by Kondo screening. A quantum phase transition point separates 
these two regimes at temperature $T = 0$. The Kondo-dominated phase is 
shown to possess soft modes, with spectral gaps much smaller 
than the Kondo temperature.  

}
  
\end{abstract}


\vskip2pc] \narrowtext

%\bigskip
%\twocolumn

In this paper we discuss a  one-dimensional two-channel
SU(N)-symmetric  Kondo 
lattice, that is, 
a 1-D lattice where spin operators are represented by generators of
the su(N) algebra and the conduction band has an additional 
two-fold degeneracy. We discuss the case where each conduction band is
half-filled. At this point due to the additional particle-hole
symmetry the total symmetry of the model is SU(N)$
\times$Z$_2\times $SU(2).

 The interest in  this model is justified by the fact that 
 the corresponding 
impurity model exhibits 
non-Fermi li\-quid (NFL) behaviour  \cite{2imp}, which also takes place  
for the $D=\infty$ two-channel Kondo lattice \cite{infd}.   
 The origin of this  NFL behaviour in the single impurity problem is
quite clear - it is caused by dispersionless modes (see 
Refs.~[\cite{emerykiv92}] ).

Our study of the half-filled lattice case demonstrates that 
 the system can exist in two phases: the RKKY dominated phase,  
where its low-energy behaviour is determined by the RKKY \cite{rkky} 
exchange interactions between local magnetic moments induced by
the conduction electrons, and the Kondo-dominated or heavy fermion state, 
where Kondo screening of the magnetic impurities takes place.  
 Both phases are separated by a quantum phase transition point at $T = 0$. 
 This agrees with the scenario for the Kondo lattice suggested by Doniach 
\cite{doniach}. The heavy fermion phase possesses soft modes; these modes 
have spectral gaps which are much smaller than the Kondo temperature ($T_K$).

Most of the papers devoted
 to the Kondo lattice use the slave-boson approach 
and the large-N approximation
 \cite{sboson}, which is more suitable for the Kondo dominated
 (disordered) phase. 
The opposite 
situation is easier to study in one dimension, where the RKKY
interaction 
has a simpler form.
 This has been done by one of the authors for the one-dimensional 
 M-channel Kondo lattice
 at half-filling \cite{rkkylimit}: when the spin sector has SU(2) symmetry.
In this case both 
  charge and orbital sectors acquire gaps, while for half-integer 
value of $ |M/2 - S| $ the spin sector 
remains gapless
 and at low energies is described by the SU$_{1}$(2) 
Wess-Zumino-Novikov-Witten model
(WZNW) \cite{wznw}. In this paper we generalize this analysis for   
SU(N) symmetry of the 
spin sector: the existence of an additional parameter, N, 
opens a  possibility to study 
  a competition between the RKKY- and the Kondo-dominated regimes.         

We consider the following Hamiltonian: 
\begin{eqnarray}
H \, & \, = \, & \,  -t \, \sum_{i;m=1,N;\alpha=1,2} c^{+}_{i,m,\alpha} 
c_{i+1,m, \alpha}   \nonumber \\ 
\, & \, & \, + \, J  \, \sum_{i;m,l=1,N;\alpha=1,2} c^{+}_{i,m,\alpha} d_{i,m} 
d^{+}_{i,l} c_{i,l,\alpha}  \,  + H.c. 
\end{eqnarray}
Here $i$ is the lattice site index, $m (= 1, ... N)$ numbers components of 
 the SU(N) spin, while 
$\alpha$ is the channel index of the conduction electrons. $c^{+}_{i,m,\alpha}$
 is the creation o\-pe\-ra\-tor for 
conduction electrons in a state characterized by quantum numbers $i,m,\alpha$, 
while $d^{+}_{i,m}$ corresponds to the local moment states. The
Hamiltonian is supplemented by the constraint
\begin{equation}
\sum_{m=1,N} d^{+}_{i,m} d_{i,m} = q,
\end{equation}
where the number $q$ represents the filling of the local orbitals. It 
remains  finite in the limit $N \longrightarrow \infty$.

The calculations for the  RKKY regime are similar to the ones for the
 SU(2)-symmetric case done in \cite{rkkylimit}. The mean field 
 gap in the charge sector is $\Delta \sim J$, 
while in the Kondo regime it is $T_{K} \sim D e^{\frac{-2 \pi}{N J 
 \rho(\epsilon_F)}}$, being $D$ 
the conduction band bandwidth and $\rho(\epsilon_F) \sim D^{-1}$ 
 the density of states at the Fermi level. 
For large $N$, the latter gap is the largest one.

The appearance of the gap leads to the change of the ground state
energy;  in the RKKY-phase we have  
\begin{equation} 
E_{RKKY} = -\frac{N \rho(\epsilon_F)\Delta^{2} }{\pi}
 ln\frac{D}{\Delta} \approx  - \frac{N\rho(\epsilon_F)J^2}{\pi} 
\ln\frac{D}{J}
\end{equation}
and in the Kondo regime it is:
\begin{equation}
-\frac{N T_{K} }{4 \pi} \, = \, \frac{N}{4 \pi} e^{\frac{-2 \pi}{N
J\rho(\epsilon_F)}}
\end{equation} 
 The critical value of $N$ at which both ground state energies 
  become  comparable is colossal
 for any realistic $J$:
\begin{equation}
 N_{c} \sim - \frac{\pi}{J\rho(\epsilon_F)} ln^{-1}[1/J\rho(\epsilon_F)]
\end{equation}
 but it does not matter since we are going to study universal properties
of the spectrum.

Let us now discuss the properties of the system on both sides of the
transition. 
Deep in the RKKY phase the adequate des\-crip\-tion of the spin sector in the
 continuum limit  is given by the $\frac{U(N)}{U(N/2) \otimes U(N/2)}$
non-linear 
sigma model (NL$\sigma$) with the topological term
\cite{read}. In the limit $N = 2$ this becomes the $SU(2)/U(1)$
model with the topological term obtained for this case in
Ref.~[\cite{rkkylimit}].
In the presence of the topological term the model 
is likely to be massless. The transmutation scale (that is the energy
scale at which the model undergoes a crossover 
 to the strong coupling regime) 
of this model is:
\begin{eqnarray}
\Delta_{ sp } \,   \, \sim \,   \, J \, e^{\frac{- 2 \pi}{N g}}, \qquad   
 g \, \sim \rho(\epsilon_F)J \sqrt{ln (D/J)}. 
\label{eq2}
\end{eqnarray}

On the other hand, in the Kondo limit the spin sector most certainly 
 has a gap (we shall see that it is much smaller than  $T_{K}$). Since
 the NL$\sigma$ model is critical, its
 renormalization group beta function \cite{betaf} has a zero at some 
 finite value of $g = \rho(\epsilon_F)JN$ (see Fig.1). 
If there is no 1st order phase transition between the two regimes, the
 beta function can be continued into the Kondo regime where the
 spectrum has a gap. For such behaviour to take place, one needs to
 have a beta function which does not cross the horizontal axis but
 just touches it. 

\vspace{0.2cm}

\begin{figure}
\epsfysize = 5cm
\begin{center}
\leavevmode
\epsffile{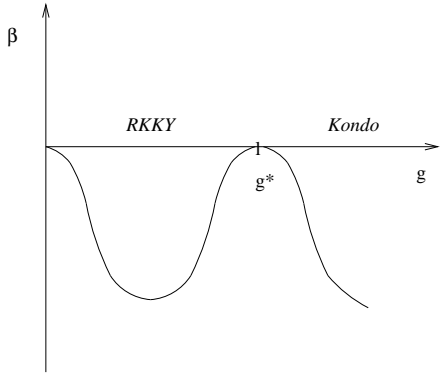}
\end{center}
\caption{ Schematic plot of the beta function as a 
 function of the bare coupling.}
\label{fig:betafunction}
\end{figure}

\vspace{0.4cm}
   
This behaviour of the $\beta$ function corresponds to the presence of 
 a marginal operator at  the  critical point. 

We will now discuss the Kondo-dominated (large $N$) regime in detail. 
 The Lagrangian density can be written as follows:
\begin{eqnarray}
L \, & \, = \, & \,   \sum_{i,j;m=1,N;\alpha=1,2} c^{+}_{i,m,\alpha}
 [ \partial_{\tau} \delta_{i,j} - \hat{\epsilon}_{i,j} ]  c_{j,m,\alpha}
 \, \nonumber\\ & & +  
 \, \sum_{i;m=1,N} d^{+}_{i,m} \partial_{\tau}\, d_{i,m}
\nonumber \\
\, &  & \, + \, \frac{J}{N} \, \sum_{i;m,l=1,N;\alpha=1,2} c^{+}_{i,m,\alpha}
d_{i,m} d^{+}_{i,l}
c_{i,l, \alpha}  \, \nonumber \\ \, &  & \, + \, i \lambda 
  \, \sum_{i;m=1,N} ( d^{+}_{i,m} d_{i,m} - q ) \, + H.c.   
\end{eqnarray}
In this expression  $\lambda$ is a Lagrange 
multiplier introduced to enforce the local constraint on  
 occupation number of the localized orbitals. 

We next decouple the interaction term employing the Hubbard-Stratonovich 
transformation:
\begin{eqnarray}
 \frac{J}{N} \, \sum_{i;m,l=1,N;\alpha=1,2} c^{+}_{i,m,\alpha}
d_{i,m} d^{+}_{i,l}
c_{i,l, \alpha}\, \,\longrightarrow \nonumber \\  \,
 \, \frac{N V^{+}_{i,\alpha} V_{i,\alpha}}{2 J}
 \, + \,    
\sum_{i;m=1,N;\alpha=1,2} [V_{i,\alpha} c^{+}_{i,m,\alpha} d_{i,m} + H.c.],
\end{eqnarray}
where $V_{i,\alpha} = \mid V \mid z_{i,\alpha}$ 
and $ \sum_{\alpha=1,2} z^{*}_{i,\alpha} z_{i,\alpha} = 1 $.
We now apply the following further transformation for 
 the conduction band orbitals:
\begin{eqnarray}
c_{i,m,\alpha} \, = \, z_{i,\alpha} \, \chi_{i,m} + 
\epsilon_{\alpha,\beta}  \, z^{*}_{i,\beta} \, \eta_{i,m},
\end{eqnarray}
which can be cast also in the following matricial form:
\begin{equation}
\left( \begin{array}{l}
{ c_{i,m,1} } \\ \\
{ c_{i,m,2} } \end{array} \right) \, = \, \left(
\begin{array}{ll}
{ z_{i,1} } & { z^{*}_{i,2} }  \\ \\
 { z_{i,2}} & { -z^{*}_{i,1}   }
\end{array}
\right) \left(
\begin{array}{l}
{ \chi_{i,m} } \\ \\
{ \eta_{i,m} }
\end{array}
\right)
\nonumber \\
\, = \, 
- \hat{g}_{i} \left(
\begin{array}{l}
{ \chi_{i,m} } \\ \\
{ \eta_{i,m} }
\end{array} \right)
\end{equation}
Here $ \hat{g} $ is a matrix belonging to SU(2). 

One can imagine two simple-minded mean field confi\-gu\-rations now:

a) the uniform one, in which the  $ \hat{g} $ matrices on different sites are equal;

b) the case $  z_{2i+1,\alpha} \, = \, 
\epsilon_{\alpha,\beta}  \, z^{*}_{2i,\beta} \,$, which in matrix form would read:
\begin{equation}
\hat{g}_{2i+1}   
\, = \, 
- \hat{g}_{2i} 
 \left(
\begin{array}{ll}
{ 0 } & { -1}  \\ \\
 { 1} & { 0 }
\end{array}
\right).
\end{equation}

In the first case, in mean field approximation only one 
 conduction band channel ($\chi$-chain) hybridizes with the local orbitals. 
 Meanwhile, in case (b) both chains are similarly hybridized (see Fig.2).

The spectrum in case (b) is determined by the following eigenvalue equation:
\begin{equation}
\omega \, ( w^{2} - 4 \, \cos^{2}(k/2) \, V^{2} ) \,  = \, 0, 
\end{equation}
so that there is a solution $\omega = 0$. The appearance of such solution 
indicates that this configuration is energetically unstable. This conclusion 
is supported by the following calculations.

\vspace{0.2cm}

\begin{figure}
\epsfysize = 10cm
\begin{center}
\leavevmode
\epsffile{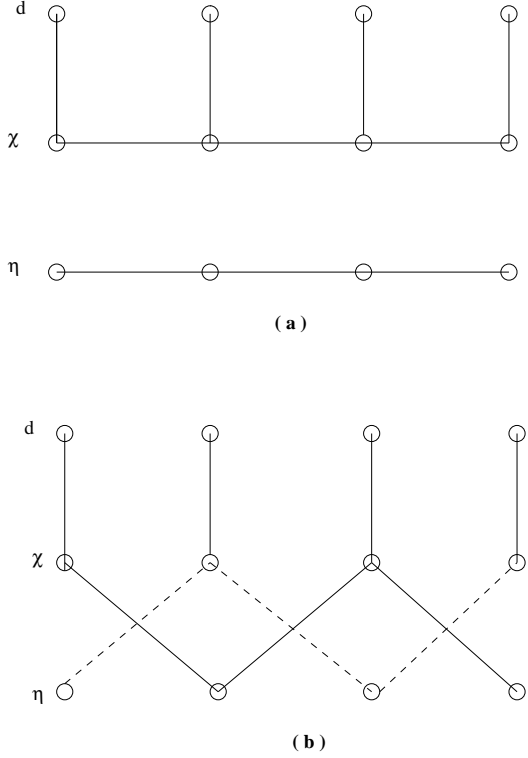}
\end{center}
\caption{ Pictorial view of the hybridization of the chains. Cases (a) and (b) 
as described in text.}
\label{fig:hybridization}
\end{figure}

\vspace{0.4cm}

%The spectrum in case (b) is determined by the following eigenvalue equation:
%\begin{equation}
%\omega \, ( w^{2} - 4 \, \cos^{2}(k/2) \, V^{2} ) \,  = \, 0, 
%\end{equation}
%so that there is a solution $\omega = 0$. The appearance of such solution 
%indicates that this configuration is energetically unstable. This conclusion 
%is supported by the following calculations.

The self-consistency equation determining $V$ is:
\begin{equation}
V \, = \, N J <\chi_{i,m} d^{+}_{i,m}>,
\end{equation}
which, upon introducing the one-dimensional tight-binding 
 dispersion relation for the conduction bands, 
yields the following expressions for $V$, respectively, in the two cases:
\begin{eqnarray}
1 \, = \, \frac{ N J}{2 \pi} \int^{\pi/2}_{-\pi/2} \frac{dk}{\sqrt{\sin^{2}k 
+ V_{(a)}^{2}}}, \nonumber \\
 1 \, = \, \frac{ N J}{2 \pi} \int^{\pi/2}_{0} \frac{dk}{\sqrt{\sin^{2}k 
+ \frac{V_{(b)}^{2}}{4}}}. 
\end{eqnarray}
One can see that $V_{(a)}$ for the uniform mean field solution is larger 
than $V_{(b)}$ ( by a factor of $ e^{\frac{\pi}{NJ}} $ ), thus the ground 
state 
energy is likely to be lower in the uniform solution as we now proceed to 
show. 

The ground state energy is determined by the derivative of
 the free energy with respect to the parameter $V$:
\begin{equation}
\frac{\partial {\it F}}{\partial V} \, = \, \frac{V}{J} \, + \, 
\frac{N}{2} \left( < \chi_{i,m} d^{+}_{i,m} > \, + \, 
 < d_{i,m} \chi^{+}_{i,m}  > \right),
\end{equation}  
which in the uniform case leads to:
\begin{equation}
{\it F} \, = \, \frac{1}{2} \left( \frac{V^{2}_{(a)}}{J} - \frac{N V^{2}_{(a)}}{2 \pi} 
\ln \frac{e}{V^{2}_{(a)}} \right)
\end{equation}
At the minimum 
 this leads to $ {\it F}^{(a)}_{0} \, = \, - \frac{N V^{2}_{(a)}}{4 \pi}$. Similarly, 
one gets for the other mean field solution:$ {\it F}^{(b)}_{0} \, = 
\, - \frac{N V^{2}_{(b)}}{8 \pi} $, which has higher energy than the uniform 
mean field solution.

Let us now consider fluctuations around the uniform mean-field solution.   
We write the $ \hat{g} $ matrix in terms of two slow fields, as follows:
\begin{equation}
\hat{g}_{n} \, = \, G(x) \, \left( \sqrt{ 1 - \vec{l}^{2} (x) } + 
i (-1)^{n} \vec{\sigma} \cdot 
\vec{l} (x) \right) ,
\end{equation} 
where $ \mid \vec{l} (x) \mid << 1 $ and $ G(x) $ belongs to SU(2).

 Neglecting fluctuations in $V$ we get 
the following action on the lattice:
\begin{eqnarray}
S \,& = &\, \Psi^{+}_{n} ( \partial_{\tau} + G^{+} \partial_{\tau} G ) \Psi_{n} +  
\Psi^{+}_{n} \hat{\epsilon} [i \partial_x +  a \partial_{x} G^{+} G ] \Psi_{n} \nonumber \\ 
\, & & \, + V ( \chi^{+}_{n,m} d_{n,m} + c. c ) + d^{+}_{n,m}  
\partial_{\tau} d_{n,m}  \nonumber \\ \, & & \, 
+ i \lambda(x) (  d^{+}_{n,m} d_{n,m} - \frac{N}{2} )
\Psi^{+}_{n+1} ( - 2 l^{2} ) \Psi_{n} \nonumber \\ \, & & \, 
 + i (-1)^{n} \Psi^{+}_{n} \vec{\sigma} \cdot 
\vec{l} (x) \Psi_{n}; \qquad
\Psi_{n} \, = \, 
\left(
\begin{array}{l}
{ \chi_{n,m} } \\ \\
{ \eta_{n,m} }
\end{array}
\right).
\end{eqnarray}

Let us now decompose the  unhybridized fermion into
 slowly varying components,   
\begin{equation}
\eta_{n,m} = R(x) (i)^n + L(x) (-i)^n,
\end{equation}
where $x = na$ ( $a$ is the lattice parameter), 
substitute this into the above effective action 
and  integrate  over fast fermionic fields $\chi, d$ with energies 
above the gap: $ \sim \frac{V^{2}}{D} = T_{K}$, and the field $l$.
 After the integration 
 we get
the following effective action for the flavour and the spin sectors:
\begin{eqnarray}
S & = & S[PCF] + S[GN], \\
S[PCF] & = & \frac{N}{2 \pi} Tr( \partial_{\tau}G^{+}\partial_{\tau}G ) \, 
 + \, 
\frac{N}{2 \pi} v^{2} Tr( \partial_{x}G^{+}\partial_{x}G ), \,  \label{G} \\
S[GN] & = & \bar\psi_m\gamma_{\mu}\partial_{\mu}\psi_m - \frac{\pi}{32N}
(\bar\psi_m\psi_m)^2, \label{GN}
\end{eqnarray}  
where $v \sim v_{F}$ and the spinor $\psi$ incorporates the 
$R,L$ fields. Both effective actions have $T_K$ as their ultraviolet
cut-off. 

The action for $G$ must vanish in the limit $V \longrightarrow 0$. 
 This will happen through the shrinking of the cutoff:  
$ T_{K} \longrightarrow 0$.

 The action in Eq.(\ref{G})  corresponds to the Principal Chiral Field 
 model on the SU(2) group. It is exactly solvable \cite{pw83}; 
the excitations have a spectral gap
\begin{equation}
\Delta_G \sim T_Ke^{- N}
\end{equation}
 Fermionic model (\ref{GN}), describing the spin sector,
 is the O(2N)-symmetric Gross-Neveu
 model \cite{gneveu} with the bare coupling constant $g = \pi/16N $.
 It is also exactly solvable and 
all excitations have gaps \cite{zamolodchikov}. The mass scale  is 
\begin{equation}
\Delta_{\psi} \sim T_K e^{-\frac{\pi}{g(N-2)}} \sim T_K e^{-16}  
\end{equation}
This gap is numerically small despite the fact that it does not
contain $N$. 

Thus, we see that the heavy fermion phase contains two modes, the spectra 
of which lie well inside of the single-particle  gap. One mode is magnetic 
 ( the one described by the Gross-Neveu model) and the other is not.

\vspace{0.2cm}

The present work was supported by a Joint Collaboration grant awarded by
 the British Council and Fundaci\'on Antorchas. The authors  
acknowledge the hospitality of their institutions throughout 
their respective stays abroad.

%\newpage

%\begin{figure}
%\caption{ Beta function as a function of the bare coupling.}
%\label{fig:betafunction}
%\end{figure}
%\begin{figure}
%\caption{ Pictorial view of the hybridization of the chains. 
%Cases (a) and (b) as described in text.}
%\label{fig:hybridization}
%\end{figure}


\begin{references}

\bibitem[*]{byline}Member of the Carrera del Investigador 
Cient\'{\i}fico of CONICET (Consejo Nacional de Investigaciones 
Cient\'{\i}ficas y T\'ecnicas, Argentina).

\vspace{0.2cm}

\bibitem{2imp}A. M. Tsvelik and P. B. Wiegmann, Z. f\"ur Phys. B{\bf 54},
201 (1984); N. Andrei and C. Destri, Phys. Rev. Lett. {\bf 52}, 364
(1984);
  A. M. Tsvelik, J. Phys. C{\bf 17}, 2299 (1984); 
 A. W. W. Ludwig and I. Affleck, Phys. Rev. Lett. {\bf 57}, 
3160(1991).
\bibitem{infd}M. Jarrell {\it et al.}, Phys. Rev. Lett. {\bf 77}, 1612(1996);
F. B. Anders, M. Jarrell and D. L. Cox,  
 Phys. Rev. Lett. {\bf 78}, 2000(1997).
\bibitem{emerykiv92}V. Emery and S. Kivelson, Phys. Rev. B{\bf 47}, 10 812
(1992); P. Coleman, L. Ioffe and A. M. Tsvelik, Phys. Rev. B{\bf 52}, 6611
(1995). 
\bibitem{rkky}M. A. Ruderman and C. Kittel, Phys. Rev. {\bf 96}, 99 (1954).
\bibitem{doniach} S. Doniach, Physica B {\bf 91}, 231 (1977). 
\bibitem{sboson}P. Coleman, Phys. Rev. B{\bf 35}, 5072(1987).
\bibitem{rkkylimit}A. M. Tsvelik, Phys. Rev. Lett. {\bf 72}, 1048(1994).
\bibitem{wznw} S. Novikov, Usp. Math. Nauk. {\bf 37}, 3 (1982); 
 E. Witten, Comm. Math. Phys. {\bf 92}, 455 (1984);  
 A. M. Polyakov and P. B. Wiegmann,
Phys. Lett. B{\bf 141}, 223 (1984).
\bibitem{read} N. Read and S. Sachdev, Nucl. Phys. B{\bf 316}, 609 (1989). 
\bibitem{betaf} M. Gell-Mann and F. E. Low, Phys. Rev. {\bf 95}, 1300 (1954);
see also e.g. ``Quantum Field Theory and Critical Phenomena'', by 
J. Zinn-Justin, Oxford University Press (1993).
\bibitem{pw83} A. M. Polyakov and P. B. Wiegmann, Phys. 
Lett. B{\bf 131}, 121 (1983)
\bibitem{gneveu} D. J. Gross and A. Neveu, Phys. Rev. D{\bf 10}, 3235 (1974).  
\bibitem{zamolodchikov} A. B. Zamolodchikov and Al. B. Zamolodchikov,
Ann. Phys. (N.Y.) {\bf 120}, 253 (1979); Nucl. Phys. B{\bf 133}, 525 (1978);
 P. Forg\'acs, F. Niedermayer and P. Weisz, Nucl. Phys. B{\bf 367}, 123 
 (1991).  


\end{references}
\end{document}